\documentclass{aa}     
\usepackage{epsfig}
\def\kms{km~s$^{-1}$}
\def\deg{$^\circ $}
\def\gal{3C~171 }
\def\hb{H$\beta$ }
\def\oiii{[OIII] }
\begin{document}
\thesaurus{11.01.2, 11.09.1: 3C 171, 11.09.4, 11.11.1, 11.17.2}

\title{Integral field spectroscopy of the radio galaxy \gal.
\thanks{Based on observations performed at the Canada France Hawaii
Telescope}}

\author {I.~M\'arquez\inst{1} 
\and 
E.~P\'econtal\inst{2} 
\and 
F.~Durret\inst{3,4}
\and 
P.~Petitjean\inst{3,4}}

\offprints{isabel@iaa.es }
\institute{
Instituto de Astrof\'\i sica de Andaluc\'\i a (CSIC),
Apartado 3004 , E-18080 Granada, Spain
\and
Observatoire de Lyon, 9 avenue Charles Andr\'e, F-69561 St Genis Laval
\and
Institut d'Astrophysique de Paris, CNRS, 98bis Bd Arago, F-75014 Paris, France
\and
DAEC, Observatoire de Paris, CNRS (UA 173), Universit\'e Paris VII, 
F-92195 Meudon Cedex, France
}
\date{Received, 2000; accepted,}

\maketitle

\begin{abstract}

We have performed integral field spectroscopy of the radio galaxy \gal
(redshift z=0.238) with the TIGER instrument at the Canada France
Hawaii telescope in the H$\beta$-[OIII]4959-5007 wavelength region. We
present the reconstructed \hb and [OIII] images and compare them to
the HST and radio maps.  We discuss the variations of the [OIII]/\hb
line ratio throughout the nebulosity. We also analyze the velocity
field in detail, in particular the presence of several components.  We
find that the kinematics derived with emission lines in the central
region (inside 1 arcsec) are compatible with a disk-like rotation of
low amplitude (50 km/s). The continuum surface brightness profile
follows an $r^{1/4}$ law, suggesting that the underlying galaxy is an
elliptical with an effective radius of 15 kpc.

We have fit two components in the region centered 2.7 arcsec to the
West and of extension 3 arcsec$^2$. We find that the blueshifted
component is an extension of the central part, whereas the second one
is redshifted by 600 km/s. In both components, line ratios and FWHM
are compatible with the presence of shocks induced by jet-cloud
interactions.

\end{abstract}

\section{Introduction}

A number of radio galaxies and quasars at various redshifts exhibit
line emission from ionized gas up to several tens or even hundreds of
kiloparcsecs from the nucleus. The existence of such gaseous envelopes
is most probably linked to the formation of galaxies and active
nuclei. Conversely, the existence of an active nucleus influences the
physical conditions of the gas, its kinematics, and star formation in
the host galaxy. However, illumination of the gas by the ultraviolet
radiation emitted by the active nucleus is probably not the only
mechanism responsible for the ionization of the gas (Clark et
al. 1998, hereafter C98, Villar-Mart\'\i n et al. 1999, Tadhunter et
al. 2000); the strong link between optical emission line and radio
properties suggests interactions between the gas and the
radio-emitting plasma, possibly because of shock heating and
subsequent ionization of the gas.

Radio-loud quasars at low redshift are well adapted to the study of
the interactions between the gas and radiation from the active nucleus
and/or the radio plasma.  This indeed becomes much more difficult for
more distant objects due to their smaller spatial extent and surface
brightness dimming. 
Integral field spectroscopy is well suited for this purpose, since,
contrary to long-slit spectroscopy, it allows to fully map the
velocity field in one exposure, providing the size of the ionized nebulosity
is compatible with the instrument field. Such a study was done for
example for three quasars with redshifts between 0.268 and 0.370 by
Durret et al. (1994), for a 0.734 redshift quasar by Crawford \&
Vanderriest (1997), for 3C~48 by Chatzichristou et al. (1999), for
four ultraluminous IRAS galaxies by Wilman et al. (1999) and for six
radio-loud quasars with redshifts between 0.26 and 0.60 by Crawford \&
Vanderriest (2000).

\gal is a radio-galaxy at a redshift z=0.238 (corresponding to a scale
of 3.2 kpc/arcsec for H$_0=75$ km s$^{-1}$ Mpc$^{-1}$ and
q$_0$=0.5). It is associated with an optical emission-line region
extending over 6 arcsec on either side of the nucleus (Heckman et
al. 1984). These authors found large velocity gradients and striking
similarities between the ionized gas and radio distributions. The
radio emission itself has two bright hot spots East and West of the
nucleus, and fainter emission extending perpendicularly to the radio
axis, North and South of these hot spots.  The correspondence between
the [OIII] and radio emissions was confirmed by Baum et al. (1988) and
by Blundell (1996). HST spectroscopy by Hutchings et al. (1998) has
revealed outward mo\-tions of the gas at several hundred km/s in the
very inner region. A line-free image of the host galaxy of \gal shows
a moderate elongation in the North-South direction (Baum et
al. 1988). Using long slit spectra along the radio axis (PA=102) C98
have shown evidence for shocks induced by jet-cloud interactions:
close radio/optical association, ionization minima almost coincident
with both radio hot spots, high velocity line splitting spatially
associated with the two inner hot spots, large line widths in the
external gas and anticorrelation between line width and ionization
state in the external gas.

In this paper we present 2D spectroscopic data on 3C~171, which allow us
to spatially analyse line ratios and kinematics and, used together
with radio images, to interpret them in terms of an important
contribution of shocks in the regions close to radio hot spots, as in
the scenario reported by C98.  The data are presented
in Section \ref{data}, the morphology and excitation of the ionized
gas are studied in Section \ref{morphology}, the kinematics in Section
\ref{kinematics} and the discussion and conclusions in Sections
\ref{discussion} and \ref{conclusion}.

\section{The data }\label{data}

\gal was observed on February 7-9, 1995 during a total exposure time
of 4 hours and 13 minutes with the TIGER instrument at the
3.60m Canada France Hawaii telescope. Individual exposures were offset
from one another in the East-West direction in order to cover
completely the object.  Airmasses were between 1.2 and 1.3. The grism
used was R300, giving a spectral resolution of 6.9~\AA\ FWHM in a
spectral interval containing the HeII4686\footnote{Note that this line
is close to the filter edge and therefore the measured fluxes can be
highly underestimated.}, \hb and [OIII]4959-5007 lines, corresponding
to a velocity resolution of 350 \kms\ at 6200~\AA. The spatial sampling
was 0.39 arcsec and each individual field about 7$\times$7 arcsec$^2$;
the seeing was 0.9 arcsec FWHM. The total spatial
coverage is about 9$\times$4 arcsec$^2$.  References for a full
description of the TIGER instrument and of the data reduction methods
for this kind of object can be found in Durret et al. (1994).

The first steps of the data reduction (from spectra extraction to
wavelength calibration) strongly depend on the instrument optics and
were thus performed with the TIGER data processing software.  The
following stages (cosmic ray removal, photometric calibration,
exposure merging, image reconstruction) were achieved using the XOasis
data reduction package, dedicated to the CFHT integral field
spectrograph OASIS. The algorithms of this new package are indeed more
powerful than the previous ones, especially for cosmic removal and
exposure merging. The reader can find a complete description of this
package in the XOasis user manual at
http://www-obs.univ-lyon1.fr/$\sim$oasis/reduc/reduc\_tiger\_frames.html.

We also make use of the HST image (two exposures of 300s taken by the
WFPC2 with the red F702W filter retrieved from the archive, realigned
and summed) published by Koff et al. (1996). 

We have retrieved the radio image published by Heckman et al. (1984)
from the 3CRR atlas (http://www.jb.man.ac.uk). Optical and radio
images have been aligned by assuming that the radio core is placed at
the centroid of the brightest nuclear concentration in optical images.

\section{Morphology and excitation of the ionized gas}\label{morphology}

\begin{figure}[tbp]
\centerline{
\psfig{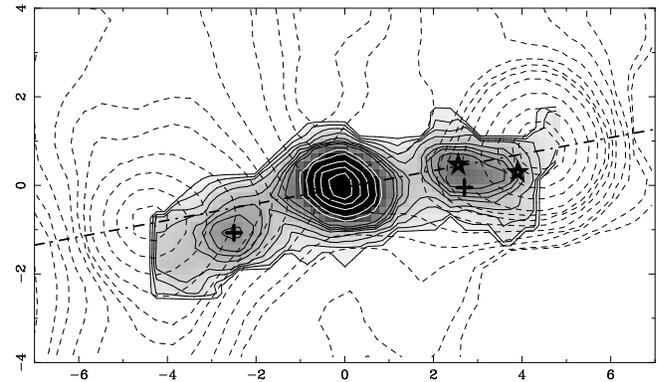}}
\caption{Contours of the [OIII] emission reconstructed from
the TIGER data, with the following values: 0.1, 0.3, 0.5, 0.7,
1, 1.2, 1.4, 1.7, 3, 4, 6, 8, and 10 ($10^{-19}$ W m$^{-2}$).
Radio contours are superimposed as dashed lines.  The
dot-dashed line joins the two radio hot spots. The positions
corresponding to the spectra presented in Figs. \ref{spec_W} and
\ref{spec_E} are shown with crosses. The positions of regions W1
and W2 are shown with stars. North is top and East to the left, and 
darker zones correspond to higher values, as in all following figures.}  
\protect\label{radioOIII}
\end{figure}

An image and contours of the [OIII] emission reconstructed from our spectra
are displayed in Fig. \ref{radioOIII}. These contours clearly reveal the
inhomogeneous structure of the nebulosity, with the ionized gas
extending roughly along an East-West direction up to the radio
hot spots, i.e. about 5 arcsec West and 4 arcsec East of the nucleus.
Strong [OIII] emission is observed in the nuclear region, as well as
in two blobs on either side of the nucleus at nuclear distances of
$\pm$2.7 arcsec. The blob 2.7 arcsec West of the nucleus will
hereafter be referred to as region W1.

In Fig. \ref{radioOIII} we also show the radio emission at 1441 MHz 
superimposed on our [OIII] image (total emission, i.e., computed by
integrating the emission in the wavelength range covered by the two
[OIII] lines). It can be seen that the peaks of the radio hot spots are
placed at the edges of the elongated structure in [OIII], about 2
arcsec further out than the [OIII] maxima.

\begin{figure}[tbp]
	\centerline{ \psfig{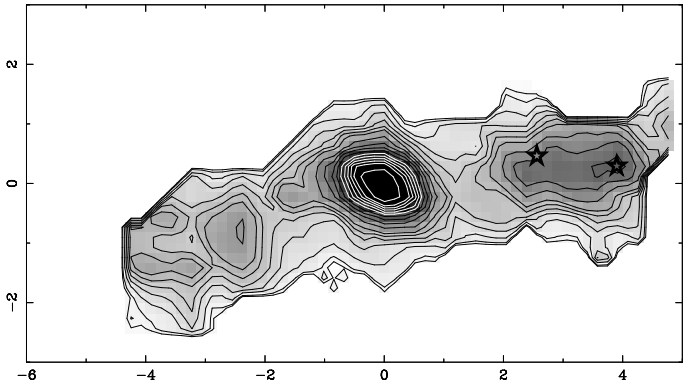}}
\caption{Contours of the H$\beta$ emission reconstructed from the
TIGER data, with the following values: 0.03, 0.05, 0.1, 0.15, 0.2,
0.25, 0.3, 0.4, 0.5, 0.6, 0.7, 0.8, 0.9, 1, 1.1, 1.2, 1.4 
($10^{-19}$ W m$^{-2}$).}
\protect\label{isophhbeta}
\end{figure}

A contour plot of the \hb emission reconstructed from our spectra is
displayed in Fig. \ref{isophhbeta}.
A region of relatively strong \hb emission is observed almost 4 arcsec
West of the nucleus, that is notably further out than region W1
observed in [OIII] West of the nucleus and therefore much closer to
the radio hot spot.  We will refer to it as region W2.

\begin{figure}[tbp]
	\centerline{ \psfig{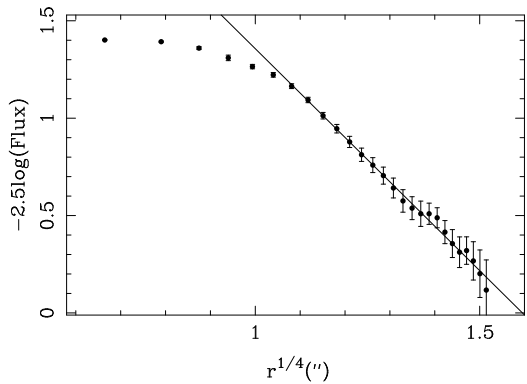}}
	\caption{Surface brightness profile obtained from the continuum 
image. The fit by a de Vaucouleurs law is displayed as a solid line.}
\protect\label{host}
\end{figure}

We have also extracted the continuum image, reconstructed by
integrating the whole wavelength range after a polynomial fit to the
continuum. It only shows the central spot which is slightly resolved
(FWHM $\approx$ 1.3 arcsec). We fit ellipses to the contours and
obtained the resulting surface brightness profile shown in
Fig. \ref{host}.  A r$^{1/4}$ law fitting results in a scalelength
(effective radius) of 4.8 arcsec, i. e., 15 kpc, a typical value for
an elliptical galaxy.

\begin{figure}[tbp]
	\centerline{ \psfig{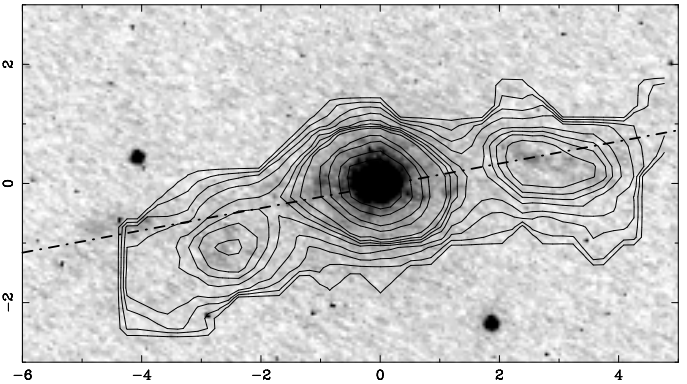}}
	\caption{[OIII] contours over HST image (filter F702W).
	The dot-dashed line joins the two radio hot spots.}
\protect\label{hst}
\end{figure}

The HST image is displayed in Fig. \ref{hst} with [OIII] contours from
our image superimposed.  The HST image was also convolved by the PSF
estimated from our direct image to create a ``smoothed HST image''
which is identical to our [OIII] image, with a bright central region
and two blobs coinciding with those in our [OIII] image. The
similarities are easily understood by taking into account the fact
that the filter used for HST observations includes the emission line
contributions from [OIII] and H$\alpha$. The emission observed in the
HST image is more extended to the East, with an almost linear feature
well aligned with the radio axis and a brightness enhancement placed at
the same position as the eastern radio hot spot. A filamentary-like
feature is also observed to the West of the nucleus, corresponding to
our West blob but this time not exactly aligned with the radio emission.

\begin{figure}[tbp]
	\centerline{ \psfig{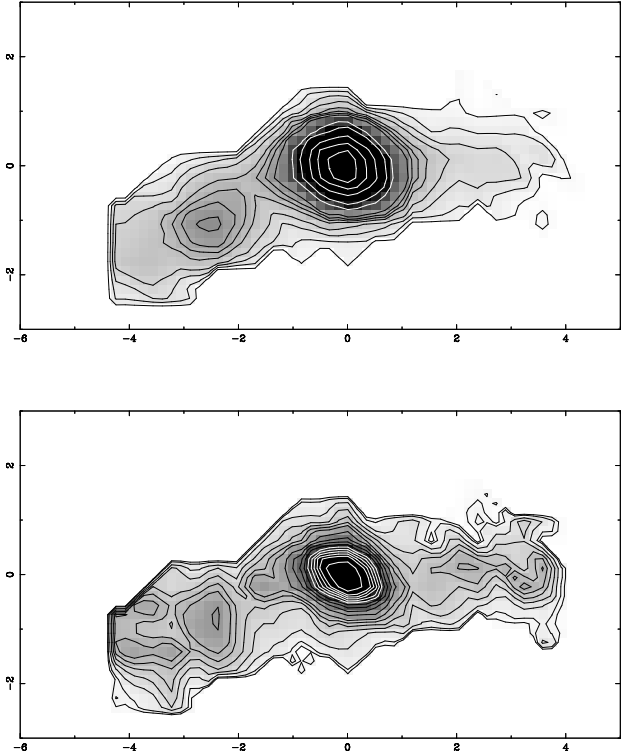}}
	\caption{Image of the blue component of \oiii (top) and \hb (bottom).
	Contour levels are 0.1, 0.3, 0.5, 0.7, 1, 1.2, 1.4, 1.7, 3, 4, 6, 8, 
	10 for \oiii and 0.03, 0.05, 0.1, 0.15, 0.2, 0.25, 0.3, 0.4, 0.5, 
	0.6, 0.7, 0.8, 0.9, 1, 1.1, 1.2, 1.4 for \hb (in units of
	$10^{-19}$ W m$^{-2}$).}
\protect\label{blue}
\end{figure}

As shown in Section \ref{kinematics}, the spectra obtained in some
regions clearly indicate line splitting, with at least two components
showing different excitations and kinematical behaviours: one about 3
arcsec to the West and another one 2.5 arcsec to the East. Two and
three components are necessary to fit the spectra in the former and
latter regions respectively; however, the signal to noise ratio (S/N)
is not high enough to allow the individual fitting of each
spectrum. These two regions are also marked in Fig. \ref{radioOIII}
with crosses. We have reconstructed two emission line images in order
to separate each of the blue and red components of [OIII] and \hb\ in
the region close to W1.  They are shown in Figs. \ref{blue} and
\ref{red}. We stress that the peak at 2.5 arcsec to the East in
Fig. \ref{blue} is located in the region where three components are
present. While the morphology of the blue component is quite similar
for [OIII] and H$\beta$, the red components are notably different, with
emission peaks shifted by about 2 arcsec, corresponding to regions W1
and W2, respectively.

\begin{figure}[tbp]
	\centerline{ \psfig{figure=marquez.f6.ps,width=5cm,angle=-90}}
	\caption{Image of the red component of \oiii (top) and \hb (bottom).
Contour levels are 0.1, 0.3, 0.5, 0.7, 1, 1.2, 1.4, 1.7, 3, 4, 6, 8, 10 
for \oiii and 0.03, 0.05, 0.1, 0.15, 0.2, 0.25, 0.3, 0.4, 0.5, 0.6, 0.7, 0.8, 
0.9, 1, 1.1, 1.2, 1.4 for \hb (in units of $10^{-19}$ W m$^{-2}$).}
\protect\label{red}
\end{figure}

\begin{figure}[tbp]
	\centerline{ \psfig{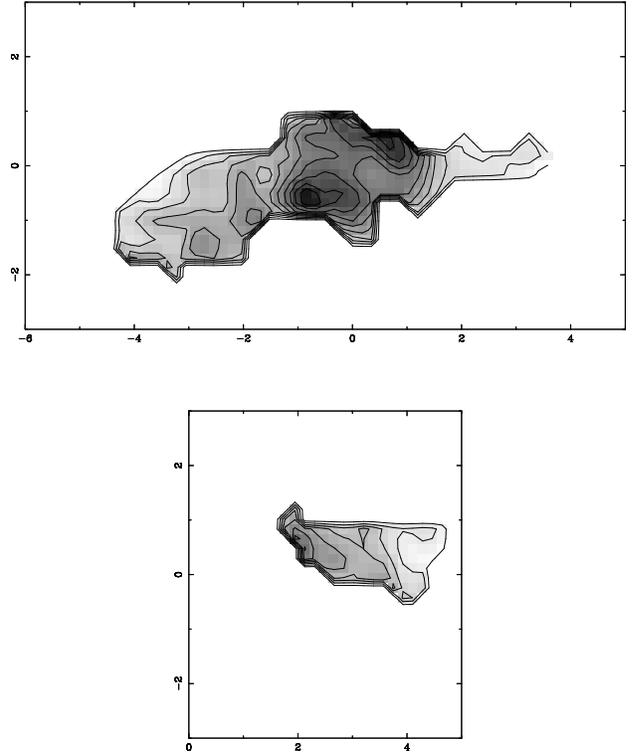}}
	\caption{[OIII]/H$\beta$ emission line ratio for the 
blue (top) and the red (bottom) component. Contours range from 1 to 10 
with a step of 1.}
\protect\label{ratio}
\end{figure}

We have computed [OIII]/\hb ratios for the blue and red components,
rejecting the spectra where the S/N for \hb is smaller than 1.5. The
resulting images are displayed in Fig.  \ref{ratio}.  It can be
noticed that, for the blue component, the regions of highest
excitation are located almost symmetrically with respect to the
nucleus at $\approx$ 1 arcsec in a direction almost perpendicular (PA=124)
to that of the elongation of the central isophotes in Fig. \ref{blue}
(PA=64). There are two local maxima, one with [OIII]/\hb= 5
corresponding to the 3$-$component region, and another one with
[OIII]/\hb= 6 placed at ($-2,-1$) arcsec. For the red component, the highest
excitation is reached in the region closest to the center of the radio
galaxy, with decreasing values towards the outskirts.

\begin{figure}[tbp]
	\centerline{ \psfig{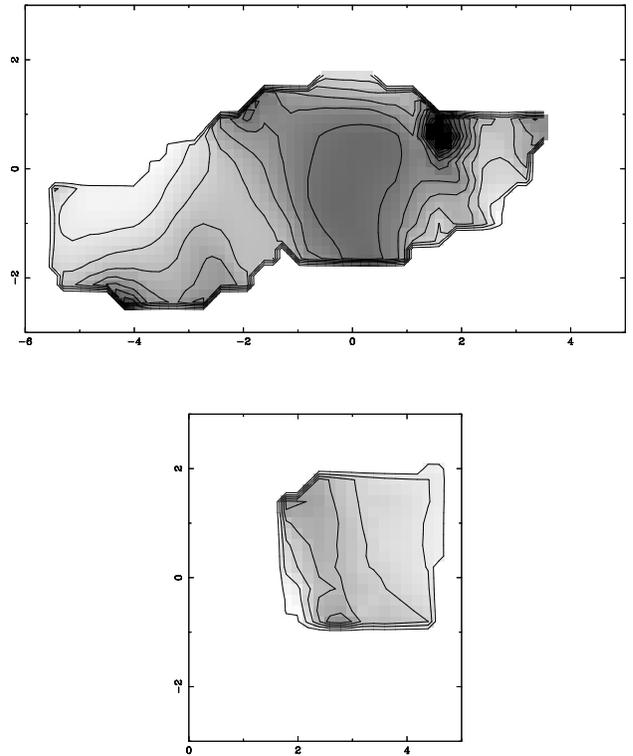}}
	\caption{[OIII]/H$\beta$ emission line ratio for the blue
	(top) and the red (bottom) component (convolved,
	S/N$>$5). Contours range from 1 to 12 with a step of 1.}
	\protect\label{ratio_ind}
\end{figure}

In order to have a high enough S/N ratio for \hb to be able to fit
[OIII] and \hb without constraining their redshifts to be the same, we
have filtered our data-cubes with a Gaussian of FWHM 1.2 arcsec. This
allowed us to get more spatially extended information on [OIII]/H$\beta$.
We then re-analysed the resulting spectra, now considering the spectra
with S/N for \hb greater than 5.  The results are shown in
Fig. \ref{ratio_ind}.  As expected, the details have been smoothed
out, and now the blue component shows a plateau in the central region,
a local peak close to the three-component region and a general trend
for decreasing ratios to the East. The maximum to the West (at
(1.5,1) arcsec) is due to the contribution of the red component in this
region (Fig. \ref{ratio_ind}, top). For the red component
(Fig. \ref{ratio_ind}, bottom), the behaviour is essentially the same
as for the non-convolved data, but extending somewhat further.

\section{Kinematics of the ionized gas}\label{kinematics}

\begin{figure}[tbp]
	\centerline{
	\psfig{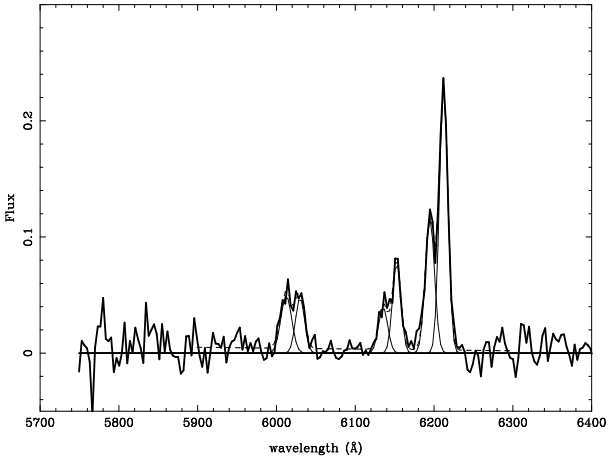}}
	\caption{Spectrum of the West region with a gaussian 
superimposed on each emission line.}
\protect\label{spec_W}
\end{figure}

\begin{figure}[tbp]
\centerline{ \psfig{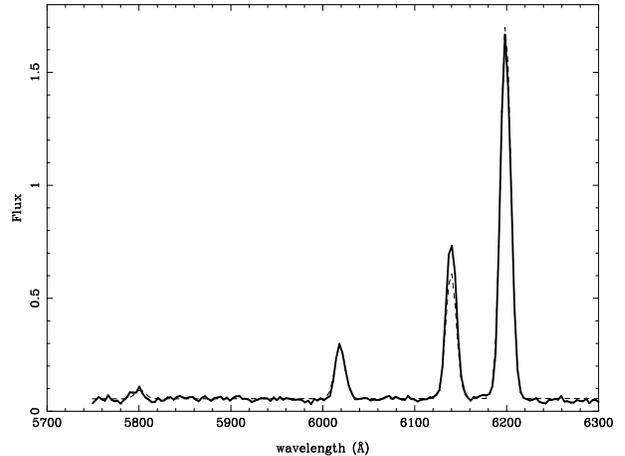}}
\caption{Spectrum of the nuclear region with a gaussian superimposed on 
each emission line.} 
\protect\label{spec_nucl}
\end{figure}

As already mentioned in section \ref{morphology}, some of the spectra
show clear signs of line splitting. These are more evident in the
region  about 3 arcsec to the West, where we could fit the emission lines
with two components. In Fig. \ref{spec_W} we show the resulting fit for
the region 2.7 arcsec West of the nucleus covering 3
arcsec$^2$. For the regions about 2.5 arcsec East of the nucleus, no
fit is possible for each individual lens, but we have extracted a
spectrum over about 1.4 arcsec$^2$ showing that at least 3 components
are present (see Fig. \ref{spec_E}). The spectrum of the nuclear
region is shown in Fig. \ref{spec_nucl}.

\begin{figure}[tbp]
	\centerline{
	\psfig{figure=marquez.f10.ps,height=8cm,width=6cm,angle=-90}}
	\caption{Spectrum of the East region with a gaussian 
superimposed on each emission line.}
\protect\label{spec_E}
\end{figure}

At difference with Hutchings et al. (1998) we actually detect the MgI
stellar absorption line feature, from which we determine the systemic
velocity to be 63126$\pm$334 \kms\ (the error bar is here the FWHM of
the instrumental spectral line broadening, and is most probably
overestimated). This value is quite close to the one derived from the
[OIII] lines in the nuclear spectrum: 63069 \kms. We will use this
value hereafter when calculating velocities relative to the
nucleus. We note that only one line of the MgI triplet is
detected. This could be due to the presence of emission lines such as
FeII which can be strong enough to fill the absorption lines (see
e.g. Boroson \& Green 1992). The low S/N ($\approx$ 6) precludes any
accurate measurement of the velocity dispersion and equivalent width
of the MgI line.  Note that this absorption feature is also present in
the nuclear low dispersion spectra of C98 as a faint dip just left of
the [NI]5199 line.

\begin{figure}[tbp]
\centerline{ \psfig{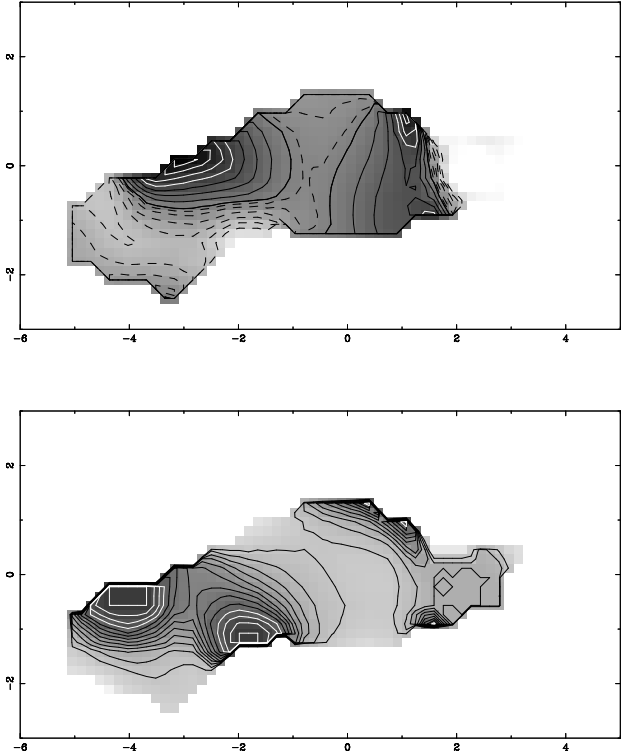}}
\caption{Velocity (top) and FWHM (bottom) distributions of the blue
component of H$\beta$. Velocities contours are given from $-$125 to 175 in
steps of 25 km/s. Negative values are represented by dashed
contours. FWHMs from 13 to 25 \AA~ with step=1 \AA, i.e., 533 to 1160 km/s.}
\protect\label{vfhbetablue}
\end{figure}

\begin{figure}[tbp]
\centerline{ \psfig{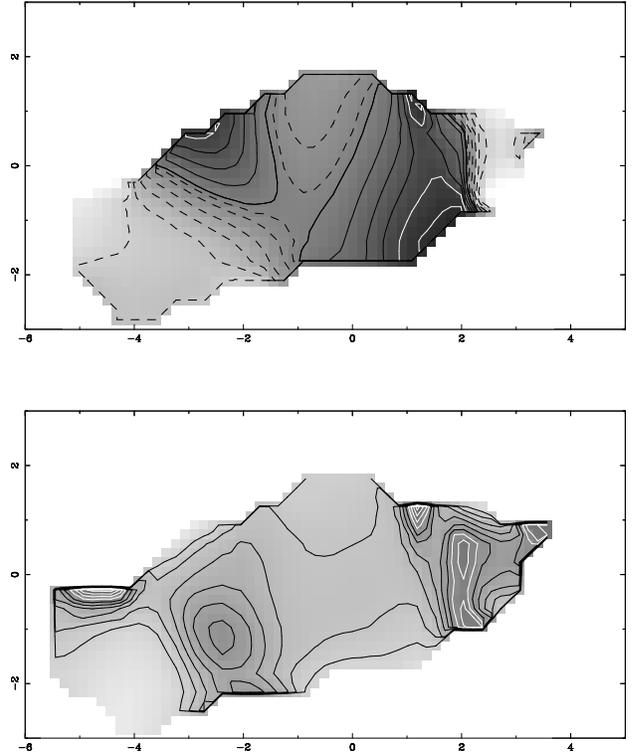}}
\caption{Velocity (top) and FWHM (bottom) distributions of the blue 
component of [OIII]. Velocities contours are given from $-$125 to 175 in
steps of 25 km/s. Negative values are represented by dashed
contours. FWHMs from 13 to 21 \AA~ with step=1 \AA, i.e., 533 to 960 km/s.} 
\protect\label{vfoiiiblue}
\end{figure}

\begin{figure}[tbp]
\centerline{ \psfig{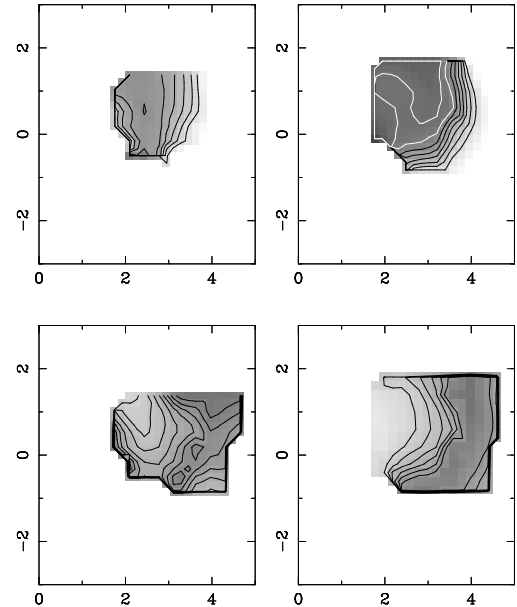}}
\caption{Velocity (top) and FWHM (bottom) distributions of the red
component of \hb (left) and \oiii (right). Velocities are contoured
from 450 to 575 km/s (step=25km/s) and FWHM from 11 to 17 \AA~(i.e.,
415 to 752 km/s once corrected from instrumental broadening. }
\protect\label{vfred}
\end{figure}

The velocity distributions have been obtained for the convolved data,
for which the S/N is high enough to allow to fit [OIII] and \hb
without constraining their redshifts to be the same.  Velocities and
FWHM distributions for the blue and red components for the two lines
are shown in Figs.  \ref{vfhbetablue}, \ref{vfoiiiblue} and \ref{vfred}.

The kinematics of the blue component, for both \hb and \oiii show a
central region of about 1 arcsec with rather well organized motions
and almost constant FHWM $\approx 13-15$ \AA\ (533 to 644 km/s once
corrected from instrumental broadening). They resemble typical
rotation, with a kinematical position angle (PA=60) in agreement with  
that of the elongations seen in the corresponding emission line maps (PA=64,
but see below for other possible interpretations). Kinematics are more
complicated for the region to the East, with velocities positive to
the North and negative to the South. We note that the FWHM reaches two
local maxima of about 25 \AA\ (1160 km/s) corresponding to the
3-component region and to the position closest to the eastern radio
hot spot. From 2 arcsec outwards to the East, velocities start
reaching negative values with FWHM somewhat higher that in the central
region. Note that velocities and FWHM maxima are higher for \hb than
for [OIII].

The red component reaches positive velocities of about 600 km/s, but
with different trends for the two lines. The peak velocity (575 km/s)
and the minimum FWHM (13 \AA) for \hb coincide with W1; velocities then
decrease almost along PA=0 on either side of the nucleus and FWHM
reach their maxima for region W2 and to the South of W2 (18 \AA,
i.e., 800 km/s). \oiii velocities are the greatest (625-600 km/s) for
the central 1 arcsec of the red component, with an almost flat
distribution around W1 and a sharp decrease to the South-West over
W2. The FWHM of \oiii slightly increases from East to West, with maximum
values (19 \AA\ or 856 km/s) around W2.

\section{Discussion}\label{discussion}

As reported by Hutchings et al. (1998), the inner optical jet (inside
$\approx$ 1 arcsec) of \gal as traced by the HST image is not aligned
with the radio axis. However, at larger sca\-les the association
between the optical emission lines and radio morphologies suggests
that the processes producing both types of emissions are closely
related. As confirmed by recent hydrodynamical simulations by Higgins
et al. (1999), a collision between an extragalactic jet and a dense
intergalactic cloud can lead to structures comparable to those
observed in 3C171.  We note that our data do not cover the tail towards the
North reported by Tadhunter et al. (2000) in an H$\alpha$ emission
line image.

The close association between radio and emission line morphologies
together with the ionization minima coincident with both radio hot
spots, the high velocity line-splitting displaced by 2 arcsec behind
the hot spots, the FWHM of about 1300 km/s and the anticorrelation
between line width and ionization state, have been considered as
evidence for shocks induced by jet-cloud interactions (C98). C98
derive these conclusions from high (about 2 \AA) and low (about 8.5
\AA) resolution long slit spectra along the radio axis (PA=122), with
1.3 arcsec spatial resolution. We use 2D spectroscopic information
with better spatial resolution (0.9 arcsec) but with spectral
resolution of 6.9 \AA. We have extracted the kinematical and emission
line values for a cut along PA=122 in order to compare with their low
resolution data. A general agreement is obtained. The only noticeable
difference is that in our data the \oiii/\hb ratio reaches a local
minimum at the center, followed by two maxima on either side at about
1 arcsec, probably due to a better spatial sampling (compare
Figs. \ref{ratio} and \ref{ratio_ind}). With respect to the presence
of various emission line components, we have only fit the regions
where a single component fit was not satisfactory (see section
\ref{kinematics}). We have only used two components for each of the
\oiii 4959, \oiii 5007 and \hb lines. C98 fit five components to
their high resolution data, so a direct comparison is not possible.

From the two component fitting we have applied to the region to the
West, we may clearly separate two regions with different kinematics
and ionization states, the red one clearly detached from the blue one,
which is associated with the central region. Note that the red
component corresponds to the linear feature on the HST image which is
not well aligned with the radio axis.

\begin{figure}[tbp]
	\centerline{ \psfig{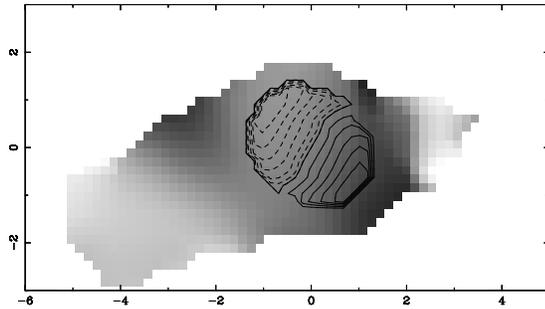}}
	\caption{Same as Fig. \ref{vfoiiiblue} (top) with a simple
	kinematic model superimposed (see text). }
	\protect\label{vit_modele}
\end{figure}

The central 2 arcsec show a velocity field that could be reminiscent
of (asymmetric) rotation with an amplitude of 50 km/s. The
kinematical position angle (PA=64) coincides with that of the central
1 arcsec elongated structure present in the HST image. \oiii and \hb
trace essentially the same kinematics. Similar rotations have been
found in some radio-loud quasars (Chatzichristou et al. 1999; Crawford
\& Vanderriest 2000). To reproduce the gas rotation we have
constructed a kinematical model representing a symmetric, inclined
disk rotation as a solid body out to r=1.1 arcsec and then a flat
rotation curve with V$_{max}$=140 km/s (from r=1.1 to 1.4 arcsec). A
non constant value of the major axis produces a better representation
of the velocity field; we have fixed the kinematical major axis at
PA=47\degr~ inside r=0.5 arcsec, then increasing linearly outwards 
to PA=67\deg. The adopted inclination is $i$=82\degr, derived from the
ellipticity of the central body.  In Fig. \ref{vit_modele} we show the
resulting isovelocity contours. In spite of the simplicity of the
kinematical model, the observed velocities are reasonably well
reproduced in this region. \oiii/\hb ratios in this central region
show two maxima with \oiii/\hb $\approx$ 10 at about 1 arcsec (3.2
kpc), at the ends of the kinematical minor axis. This result is due to
the fact \hb contours are more elongated than \oiii contours.

Based on a long slit spectrum along PA=60 with the STIS on HST,
Hutchings et al. (1998) have suggested that a central outflow in this
direction could also explain the observed kinematics, consistently
with the clumpy emission-line structures indicating outward motions of
a few $\approx$ 100 km/s within a centrally illuminated and ionized
biconical region.  For the sake of comparison we have extracted a cut
along this PA; we obtain a small amplitude (about 80 km/s), smooth
velocity distribution, where the local peaks detected by Hutchings et
al. at about $-$0.3, 0.1 and 0.5 arcsec, are absent.  Their better
spatial resolution ($\approx$ 0.1 arcsec) could explain the
differences. However, such a model requires a mechanism for
bending the jet from PA=60 to $\sim$100 over a distance of about 1
arcsec. Notice that the spatial resolution of the radio map is not
sufficient to confirm the presence of such a central jet.

The red component 3 arcsec to the West is detached by about 600 km/s
from the blue one, and this cannot be reconciled with a gravitational
origin. \oiii and \hb kinematics are slightly different, with
differences up to about 200 km/s for the two lines. We note that, with
the exception of the nuclear region, we do not achieve high enough S/N
to extract HeII 4686 fluxes that would allow to analyze HeII/\hb {\sl
versus} \oiii/\hb in the context of the models of shock $+$ precursor
of Dopita \& Sutherland (1995), in the same way as Feinstein et
al. (1999) for the radio galaxy 3C~299. The emission line FWHM
reach values of about 860 km/s in the regions where \oiii/\hb ratios are
minimal, in agreement with the hypothesis of jet-cloud interaction
processes, as decribed by C98 (see above).

Such interactions are expected to be more important in rich
environments, which seems to be the case at higher redshifts; in fact,
\gal has been proposed by C98 as an intermediate redshift prototype of
high redshift radio galaxies. However, Baum et al. (1988)
reported that \gal is a very isolated object, with the closest
possible companion at 200 kpc in projection. Moreover, from ROSAT
X-ray images, McNamara et al. (1994) concluded that \gal is not
associated with a rich cluster of galaxies (whereas they did not
exclude its association with a poor cluster or group). This also seems
to be the case for the intermediate redshift FRII radio galaxy 3C~299; 
it also shows jet-cloud interaction producing shocks that ionize the gas and
produce the radio optical alignement effect (Feinstein et al. 1999),
and is also reported to reside in a non-cluster environment
(Wan \& Daly 1996; Zirbel 1997).

\section{Conclusions}\label{conclusion}

We have mapped for the first time the extended ionized gas around \gal
in the \oiii and H$\beta$ emission lines and derived the kinematical
and physical properties. We have found that the properties of the
central region can be interpreted in terms of those of a typical ENLR
disk of radius 1 arcsec (3.2 kpc) following a low amplitude 
rotation.  The continuum surface brightness profile follows an
$r^{1/4}$ law, suggesting that the underlying galaxy is an elliptical
with an effective radius of 15 kpc.

The kinematics are much more complicated when approaching the radio
hot spots, with clear line splitting. Two components can be fit in the
West region, corresponding to an extension of the central region and
to a detached blob at about 600 km/s. Line ratios and FWHM are
compatible with the jet-cloud interaction scenario proposed by C98.

\gal is quite an isolated object, at most belonging to a poor cluster
or group, with properties resembling those of high redshift radio
galaxies. Such high redshift radio galaxies reside in much richer
environments, which are invoked to explain the origin of such
jet-cloud interaction.  In intermediate redshift radio galaxies
showing evidence for shocks produced by jet-cloud interactions as the
origin of optical-radio alignements, the mechanisms are more likely to
be related with the ambient gas, since a number of these objects
neither have nearby companions nor rich cluster environments.

\begin{acknowledgements}
We acknowledge discussions with E. Emsellem, P. Ferruit, J. Masegosa
and M. Villar-Mart\'\i n.  We are also very grateful to J. Perea for
his help setting up his SIPL graphics package at the IAP.  This work
is financed by DGICyT grants PB93-0139 and PB96-0921. Financial
support to develop the present investigation has been obtained through
the Junta de Andaluc\'{\i}a.

\end{acknowledgements}

\end{document}